\def \abc#1#2#3#4 {\reference#1, {\sl#2}, {\bf#3}, #4}
\def \blank {\lower 5pt\hbox to 0.75in{\hrulefill}}

\def \cm{~\rm{cm}}
\def \s{~\rm{s}}
\def \km{~\rm{km}}

\def \g{~\rm{g}}

\def \yr{~\rm{yr}}
\def \pc{~\rm{pc}}

\documentclass[12pt,preprint]{aastex}
\begin{document}
%\normalsize
\small

\setcounter{page}{1}
% simwin7.tex (12 Oct. 1999: WITH CORRECTIONS INSIDE, Sub. To ApJ)

\begin{center}
\bf
COLLISIONS OF FREE FLOATING PLANETS WITH \\
EVOLVED STARS IN GLOBULAR CLUSTERS
\end{center}
%\vspace*{2.0cm}

\begin{center}
Noam Soker\\
Department of Astronomy, University of Virginia \\
and \\
Department of Physics, University of Haifa at Oranim\\
Oranim, Tivon 36006, ISRAEL \\
soker@physics.technion.ac.il\\
\bigskip
Saul Rappaport \& John Fregeau \\
Physics Department, MIT \\
Cambridge, MA 02139 \\
sar@mit.edu; fregeau@mit.edu
\end{center}

%\clearpage
%$$
%$$

\begin{center}
\bf ABSTRACT
\end{center}

We estimate the rate of collisions between stars and free-floating
planets (FFPs) in globular clusters, in particular
the collision of FFPs with red giant branch (RGB) stars.
Recent dynamical simulations imply that the density of such
objects could exceed $\sim 10^6$ pc$^{-3}$ near the cores of
rich globular clusters. We show that in these clusters
$\sim 5-10 \%$ of all RGB stars near the core would suffer
a collision with a FFP, and that such a collision can spin
up the RGB star's envelope by an order of magnitude.
In turn, the higher rotation rates may lead to enhanced mass-loss
rates on the RGB, which could result in bluer horizontal
branch (HB) stars.  Hence, it is plausible that the presence
of a large population of FFPs in a globular cluster can influence
the distribution of stars on the HB of that cluster to a
detectable degree.

%\clearpage

{\it Subject headings:} globular clusters: general
--- globular clusters
--- stars: horizontal-branch
--- stars: planets

% ======================================================================
\section{INTRODUCTION}
% ======================================================================
 Recent photometric and spectral observations of young star
clusters have led to the discovery of many free-floating substellar objects,
i.e., not in orbit with a star (Mart\'in {\it et al.} 2001, and
references therein). More recently, the microlensing survey
of the globular cluster M22 (NGC 6656) has led to the
{\em highly tentative} discovery of six free floating
planets (Sahu {\it et al.} 2001).  Intuitively, one would expect
that low-mass objects, including free floating planets (hereafter
FFPs), would be expelled from a globular cluster (GC)
in a time much less than a typical age of GCs.  Specifically,
equipartition of energy between stars and planets would lead to planets
with velocities well in excess of the cluster escape speed.  However,
recent numerical simulations (Fregeau, Joshi, \&
Rasio 2001; see also Hurley \& Shara 2001) show clearly
that FFPs can survive in GCs, with a substantial fraction of the
original FFPs retained at the current epoch, and having
a velocity distribution whose rms speed is only roughly twice
that of the stars.  The survival probability increases with GCs
which have an initially higher central concentration. Fregeau et al.
(2001) have shown that globular clusters with an initial mass fraction
in FFPs of $\sim 20\%$ could evolve to the current epoch with an
FFP population which exceeds the stellar population at the cluster
center by a factor of $\sim 100$. If correct, this would
lead to the obvious conclusion that the rate of collisions
between FFPs and stars will be larger than the stellar collision
rate by a similar factor (see discussion below). Since
stellar collisions are generally non-negligible in GCs, as
evident from the presence of a blue straggler population (Shara 1999),
it is worth examining the possible influence of FFP-stellar
collisions on the observed Hertzsprung-Russell (HR) diagram of GCs.

The collision between an FFP and star will have a much lesser
effect than a collision with another star due to the fact
that a planet will add very little mass and release only a
small amount of gravitational energy. However, when entering
the envelope of a giant star, whether on the red giant branch
(RGB) or later on the asymptotic giant branch (AGB),
FFPs may deposit a substantial amount of angular momentum,
spinning-up the star by a factor of up to $\sim 100\times
(m_p/M_{\rm J})$, where $m_p$ and $M_{\rm J}$ are the masses 
of the planet and of Jupiter, respectively.
(The same comments apply to orbiting
planets that are swallowed by the expansion of their parent
RGB star; Siess \& Livio 1999; Soker \& Harpaz 2000).
The faster rotation induced by the collision may lead
to a higher mass-loss rate (Siess \& Livio 1999; Soker \& Harpaz 2000),
and since RGB stars which lose more mass
become bluer horizontal branch (HB) stars (e.g., Rood 1973; Catelan 1993;
D'Cruz {\it et al.} 1996; Brown et al. 2001), planets may play a role in
determining the distribution of stars on the HB of the HR diagram (Soker
1998a), the so-called HB morphology. Although the direct conenction between
faster rotation and mass loss is not known, rotation appears to be
the best candidate to enhance mass-loss rates in RGB stars
(R. Rood, private communication).

Motivated by the above arguments, we have carried out a study to estimate
the number of FFP-stellar collisions expected for stars that have evolved
off the main sequence. In \S2 we calculate the probability
that a star in any evolutionary phase will collide with a planet. In \S3 we
calculate the average deposited angular momentum. We summarize our main
results in \S4.

% ======================================================================
\section{COLLISION PROBABILITIES}
% ======================================================================

	The cross section for a FFP and a star to pass within a distance of
closest approach, $s$, is given by:
\begin{eqnarray}
\sigma = \pi \left[s^2 + \frac {2 s G (M+m_{p})}{v^2}  \right]  ,
\end{eqnarray}
where $M$ and $m_p$ are the masses of the star and the FFP, respectively,
and $v$ is the relative speed of the two objects when they are far apart
(see, e.g., Rappaport, Putney, \& Verbunt 1989; Di~Stefano \& Rappaport
1992). The first term in brackets is the geometrical cross
section, while the second term represents the
contribution from ``gravitational focusing''. For a star
situated in a region containing a uniform space density, $n_0$,
of FFPs, the rate at which a typical star will have an encounter
with a FFP in which the distance of closest approach is smaller than $s$,
hereafter referred to as the probability of a collision
per unit time, $\dot p(s)$, is given by:
\begin{eqnarray}
\dot p(s) \equiv \frac {dp}{dt} = \int_0^{\infty} n_{0} f(v) \sigma (v,s) dv ,
\end{eqnarray}
where we have averaged the velocity-dependent cross section over the
appropriate relative speed distribution, $f(v)$, between stars and FFPs.
If we assume that $f(v)$ can be represented by a Maxwell-Boltzmann
distribution with a 1-dimensional rms relative speed of $v_0$, then
equation (2) reduces to:
\begin{eqnarray}
\dot p(s) = 2 n_0 (2 \pi)^{1/2}
\left(s^2 v_0 + \frac { s G M}{v_0}  \right)
\end{eqnarray}
(see equation 3.4 of Di~Stefano \& Rappaport 1992), where
we have neglected the mass of the planet in comparison
with the stellar mass.

	We now assume that a collision will take place if the distance of
closest approach $s$ is smaller than the stellar radius $R$.
For stars of mass $\sim 1~M_\odot$ and radius $R \lesssim 3~R_\odot$, the
approaching planet will disintegrate due to tidal forces, while for
larger stellar radii, the planet will strike the stellar surface intact.
However, even for the case of tidal breakup, we expect the planetary debris
to strike the star if $s \lesssim R$, and thereby transfer all
of its orbital angular momentum.

In order to compute the probability of a planet-star collision, we
need to know how much time the star spends at each stellar radius interval
during its lifetime.  Since stars of mass $0.8 \lesssim M \lesssim
2~M_\odot$ follow the well-known core mass--radius and core
mass-luminosity relations (Refsdal \& Weigert 1970, 1971; Rappaport
{\it et al.} 1995; Eggleton 2001) once they have entered the
giant phase, it is straightforward to derive an approximate
analytic expression for the ``dwell time", $dt$, for a
star anywhere beyond the subgiant phase to be found
with radius between $R$ and $R+dR$:
\begin{eqnarray}
\frac {dt}{dR} \simeq F R^{-2},
\end{eqnarray}
where we estimate the constant to be $F \simeq 3 \times 10^{27} \s \cm$
(see also Webbink, Rappaport, \& Savonije 1983). We estimate that the
power-law dependence on $R$ given in eq. (4) is accurate to $\pm 0.2$
in the exponent.
We then set $s = R$ in equation (3) above, and multiply both sides by
$dt/dR$ to produce a collision probability (with a planet) per unit radius
interval of the evolving star.  The result is:
\begin{eqnarray}
\frac {dp}{dR} \simeq 2 n_0 (2 \pi)^{1/2} F
\left(v_0 + \frac {G M}{v_0 R}  \right).
\end{eqnarray}
If we now integrate equation (5), we find
\begin{eqnarray}
p \simeq 0.029
\left( \frac {n_0}{10^6 \pc^{-3}} \right)
\left[0.25
\left( \frac{R_2-R_1}{100 R_\odot} \right)
\left( \frac{v_0}{20 \km \s^{-1}}\right) +
\left( \frac{v_0}{20 \km \s^{-1}}\right)^{-1}
\ln(R_2/R_1) \right]
\end{eqnarray}
where we have taken the stellar mass to be $0.85 M_\odot$,
and normalized the 1-dimensional rms relative speed between
planets and stars to 20 km s$^{-1}$. This is the probability
that a planet--star collision will take place while the star expands from
radius $R_1$ to $R_2$.  Since equation (4) really applies to the subgiant
phase and beyond, the results given in equation (6) are most accurate for
low-mass stars with $R \gtrsim 3 R_\odot$.  Equation (6)
implies that there is a $\sim 7\%$ probability for a star to collide
with a planet sometime during the star's growth from 10 to
100 $R_\odot$.  This probability is obviously sensitive
to the normalization value for the density of planets,
$n_0 = 10^6$ pc$^{-3}$. For substantially lower planet densities the
probability becomes negligible, while for somewhat higher densities the
probability can be rather appreciable.

Repeating the same calculation for main sequence stars of
$M=0.85 M_\odot$ and $R=0.8 R_\odot$,
we find that the probability for the star to collide with a FFP
during its $10^{10} \yr$ main-sequence life is $\sim 18 \%$.
This shows that collisions of FFPs with main sequence stars
in Galactic globular clusters will not
deplete the FFP population much. Note also that the collisions
will not deposit as significant an amount of angular momentum as in
the case of an RGB star (see next section). The probability that
a HB star, with a life span of $\sim 10^8 \yr$, will suffer a FFP
collision is only $\sim 1 \%$ (for $n_0=10^6 \pc^{-3}$).

Recent numerical simulations of globular clusters by
Fregeau {\it et al.} (2001) clearly demonstrate that a
substantial fraction, $20-80\%$, of FFPs can survive to the current
epoch.  This study also shows that the density profile of FFPs
evolves rapidly during the early history of the cluster (i.e., in
the first $5 \times 10^8$ yr), and then approaches a well-defined
asymptotic structure at the current epoch with the ratio of
planets to stars increasing dramatically with radial distance
from the cluster center.  For model clusters of moderate initial
central concentrations, Fregeau et al. (2001) find that for a current
mass fraction in FFPs of $\sim 10\%$ (for the entire cluster),
the central density in planets (each of mass $0.25 M_{\rm J}$) would be
$\sim 2 \times 10^{5}$ pc$^{-3}$ at the current epoch.
However, if we consider the `top 20' non-core-collapse globular
clusters in terms of their central stellar densities (Harris 1996),
we estimate that such clusters could plausibly have central
planetary densities of $\sim 10^{6}$ pc$^{-3}$. We adopt this
somewhat optimistic normalization value for $n_0$; thus, our results
will pertain more to the richer, non-core-collapsed clusters.

% ======================================================================
\section{DEPOSITION OF ANGULAR MOMENTUM}
% ======================================================================

At the distance of closest approach, $s$, the planet's velocity and
specific angular momentum are $v=v_0[1+(R_b/s)]^{1/2}$ and
$j=sv_s$, respectively, where
\begin{eqnarray}
R_b \equiv \frac { 2 G M}{v_0^2}.
\end{eqnarray}
 The collision rate per unit interval in $s$ for a star of radius $R$
to engulf a planet with closest approach $s$ is given by $d\dot p/ds$.
 From equation (3) we find:
\begin{eqnarray}
\dot p^{\prime} \equiv \frac {d\dot p}{ds} = 2 n_0
(2 \pi)^{1/2} v_0 (2s + R_{b}/2).
\end{eqnarray}
 The average specific angular momentum per collision for a star
of radius $R$ is
\begin{eqnarray}
j_{\rm ave} = \frac {1} {\dot p(R)} \left( \int_0^R \dot p^{\prime} j ds
\right) ,
\end{eqnarray}
which, when written out, is:
\begin{eqnarray}
j_{\rm ave} = \frac{v_0}{R^2} \left(1+ \frac{R_b}{2R} \right)^{-1}
\int_0^R \left( 2 s +\frac{R_b}{2} \right)
\left(s^2 + s R_b \right)^{1/2} ds .
\end{eqnarray}
  Integration of equation (10) yields the average
specific angular momentum deposited in stars with radius $R$:
\begin{eqnarray}
\frac{j_{\rm ave} }{v_0 R }=
\left(24+12 a \right)^{-1}
\left[ \sqrt{1+a}~(16+10a-3a^2) + 3 a^3 \ln
\frac {1 +\sqrt {1+a}}{\sqrt{a}} \right],
\end{eqnarray}
where $a \equiv R_b/R$.
A simple useful approximation to equation (11) can be obtained if we note
that for our canonical values, $v_0 \lesssim 20 \km \s^{-1}$,
$M\gtrsim 0.85 M_\odot$ and $R \lesssim 100 R_\odot$, we have $a>8$.
We then carry out a Taylor-series expansion of equation (11) in the
variable $1/a$ and keep only the leading term:
\begin{eqnarray}
\left( \frac {j_{\rm ave}}{v_0 R} \right)_{a \gg 1} \simeq
\frac{2}{3} a^{1/2}  .
\end{eqnarray}

The maxium specific angular momentum a FFP can deposit into a star
with radius $R$ is obtained for $s=R$ and it is
\begin{eqnarray}
\frac{j_{\rm max}}{v_0 R }=({1+a})^{1/2} .
\end{eqnarray}

   These values should be compared with the specific angular momentum
deposited by an orbiting planet.
 Because of tidal interactions, the envelope of RGB stars will engulf
stars having an orbital separation of $r \sim 4 R$ (Soker 1998a).
 The specific angular momentum of an orbiting planet is therefore
\begin{eqnarray}
\frac{j_{\rm orb}}{v_0 R } = (2a)^{1/2}.
\end{eqnarray}

The values of angular momentum, $J$, implied by equations
(11) -- (14) as functions of the stellar radius $R$ are
plotted in Figure 1 (solid, dotted, dashed-dotted,
and dashed, lines, respectively) as $J = jM_{\rm J}/J_\odot$, where
$J_\odot$ is the present angular momentum of the Sun
$\sim 1.7\times 10^{48} \g \cm^2 \s^{-1}$. These units
facilitate direct comparison with commonly used values.
 Therefore, the values on the graph crudely
indicate the factor by which the
planets will spin up the star they collide with, with $R$ being the
radius of the star at the time of the collision.
 From the graph we see that FFPs can spin up RGB stars by a factor
of up to $\sim 50$, with an average factor, over all collisions in
all RGB stars, of $\sim 20$ (marked on the graph by a short
horizontal line marked $J_{\rm coll}$). We now derive this value.

In equation (11), an approximation of which is given
in eq. (12), we derived the average specific angular momentum
deposited in stars for which the collisions
take place when the stars have  radius $R$. However, the stars have a
continually evolving radius. Therefore, of somewhat greater interest
is the average specific angular momentum deposited in stars as they
evolve from $R_1$ to $R_2$:
\begin{eqnarray}
j_{12} =p^{-1}\int_{R_1}^{R_2} \frac{dp}{dR} j_a(R) dR,
\end{eqnarray}
where $dp/dR$ is given by equation (5) and $p$ by equation (6).
 If we approximate the expressions for $dp/dR$ and $p$ by using only the
``gravitational focusing'' portion of each one (last terms
in equations (5) and (6)), and utilize the approximate
expression (eq. 12) for $j_a$, we can
integrate equation (15) to find the following simple expression:
\begin{eqnarray}
\frac{j_{12}}{v_0 R_2 }=\frac{4}{3} a_2^{1/2}
\left[\frac{1-(R_{1}/R_{2})^{1/2}}{\ln (R_{2}/R_{1})} \right] ,
\end{eqnarray}
where $a_2 \equiv 2GM/v_0^2 R_2$.
For the typical case we are considering, the ``gravitational
focusing" term dominates, i.e., $a_2 >> 1$, the quantity
in brackets varies only between 0.30 and 0.42 for
$R_{2}/R_{1}$ ranging between 10 and 2.

We can now utilize equation (16) to estimate the average angular momentum,
$J_{\rm coll}$ that would be injected into the stellar envelope of a giant
by the time it reaches the tip of the RGB if there has been a collision
with a planet:
\begin{eqnarray}
J_{\rm coll} \simeq 0.5 a_{2}^{1/2} v_0 R_g m_p = 0.7 m_p (G M_g R_g)^{1/2} ,
\end{eqnarray}
where $M_g$ and $R_g$ are the mass and radius of the star at the
tip of the giant branch, respectively, $m_p$ is the mass of the
colliding planet, and $R_2$ in equation (16) has been set
equal to $R_g$. Note that equation (17) is independent of $v_0$.
Finally, we can estimate the factor by which colliding FFPs enhance
the angular momentum of giant envelopes over and above their
nominal angular momentum, which we take to be of order $J_\odot$:
\begin{eqnarray}
\frac{J_{\rm coll}}{J_\odot} \simeq 22
\left( \frac {m_p} {M_{\rm J}} \right)
\left( \frac {R_g} {100 R_\odot} \right)^{1/2}  ,
\end{eqnarray}
where we have taken $M_g = 0.85 M_\odot$.
The value of the leading coefficient in eq. (18) is plotted in Fig. 1
as a reference.  For giants in globular clusters the angular
momentum is likely to be factors of several times lower than $J_\odot$
(see, e.g., Sills \& Pinsonneault 2000) due to angular momentum losses
on the main sequence as well as on the giant branch. This would make
the enhancement factor expressed in eq. (18) somewhat larger.  If on
the other hand, we had normalized the results to Saturn-like planets,
the net enhancement factor would remain roughly as given by eq. (18).

% ======================================================================
\section{SUMMARY AND CONCLUSIONS}
% ======================================================================

We have shown that if the cores of rich globular clusters
have free-floating planet densities of $\sim 10^6$ pc$^{-3}$,
that $\sim 5-10\%$ of all RGB stars in the core would suffer
a collision with such an FFP. Such collisions would, on average,
increase the rotational angular momentum of the RGB star by
more than an order of magnitude (see eq. 18). We speculate
that the greatly enhanced rotation rates may lead to enhanced
mass-loss rates during the RGB phase (Siess \& Livio 1999; Soker
and Harpaz 2000).

To help quantify the importance of the collision-induced angular
velocity of the RGB stars, we compare it with the Keplerian
angular velocity at its equator.  Since most of the angular momentum of
the spun-up RGB star is in its envelope, we take $J_{RGB}=I_e
\omega$, where $\omega$ is the solid-body angular velocity
(a good assumption in the convective envelope), and
$I_e$ is the envelope's moment of inertia
$I_e=\alpha M_e R_{g}^2$, where $M_e$ is the envelope mass and $\alpha
\simeq 0.1$ (Soker \& Harpaz 2000). Thus, if we equate the
envelope angular momentum to the value of $J_{\rm coll}$ given in
equation (17), we derive the spun-up RGB angular velocity in the form
\begin{eqnarray}
\frac{\omega}{\omega_{\rm Kep}} \simeq 10^{-2}
\left( \frac {m_p} {M_{\rm J}} \right)
\left( \frac {M_e} {0.4 M_\odot} \right)^{-1}
\left( \frac {\alpha} {0.1} \right)^{-1}  ,
\end{eqnarray}
where $\omega_{\rm Kep}=(G M_g/R_{g}^3)^{1/2}$ is the Keplerian angular
velocity of an orbit on the stellar equator, and we
took $M_{g}=0.85 M_\odot$.

 Although the above value of $\omega/\omega_{\rm Kep}$ seems small,
it may actually be quite significant.  First we note that presently
the sun has $\omega/\omega_{\rm Kep} = 4.5 \times 10^{-3}$, and shows
axisymmetric rather than spherically symmetric surface activity.
This means that the magnetic field dictates the activity, including the
solar wind properties. In RGB and AGB stars, it is radiation
pressure (acting on grains) rather than magnetic activity
that dictates the wind properties.
 However, due to the strong convection in RGB and AGB envelopes
magnetic activity is expected, despite the very slow rotation.
Any increase in the slow rotation rate may significantly enhance surface
magnetic activity to a level where cool magnetic spots can be formed.
Dust formation, hence mass-loss rate, is supposedly enhanced above
these cool spots.
 The fact that most planetary nebulae have axisymmetrical rather than
spherical structure, but not all of these have binary star companions,
hints that slow rotation can indeed dictate some properties of the mass
loss process.
 Based on a crude estimate, Soker (1998b) argues that  rotation velocities
of $\omega \gtrsim 10^{-4} \omega_{\rm Kep}$, are sufficient to lead to
magnetic activity which may form cool magnetic spots on the surface of AGB
stars.
 If this holds for RGB stars, then even planets much lighter than Jupiter
may influence the mass loss process. The more mass the star loses on
the RGB the bluer the HB star it becomes.

 Hence, our main claim in the present paper is that the presence of
a large population of FFPs in a GC can lead to a potentially significant
population of blue, and extreme blue, HB stars.

%====================================================================
%====================================================================
\bigskip

{\bf ACKNOWLEDGMENTS:}
We are grateful to Bob Rood and Eric Pfahl for helpful
discussions.  We also thank the referee, Jarrod Hurley,
for extremely valuable comments on the text.
This research was supported in part by grants from the
US-Israel Binational Science Foundation, and NASA under
its Astrophysics Theory Program: Grants NAG5-4057 and NAG5-8368.
% ===================================================================

% ===================================================================

% Fixing the figure size (from Catelan):
%\parbox{3in}{\epsfxsize=3.25in \epsfysize=3.0in \epsfbox{Figure1.eps}}
%\vskip 0.1in
%\centerline{\parbox{3.5in}{\footnotesize  {\sc Fig.~1.---}
%       This is my figure caption.
%\label{fig1}
%       }}

\begin{figure}
\figurenum{1}
\plotone{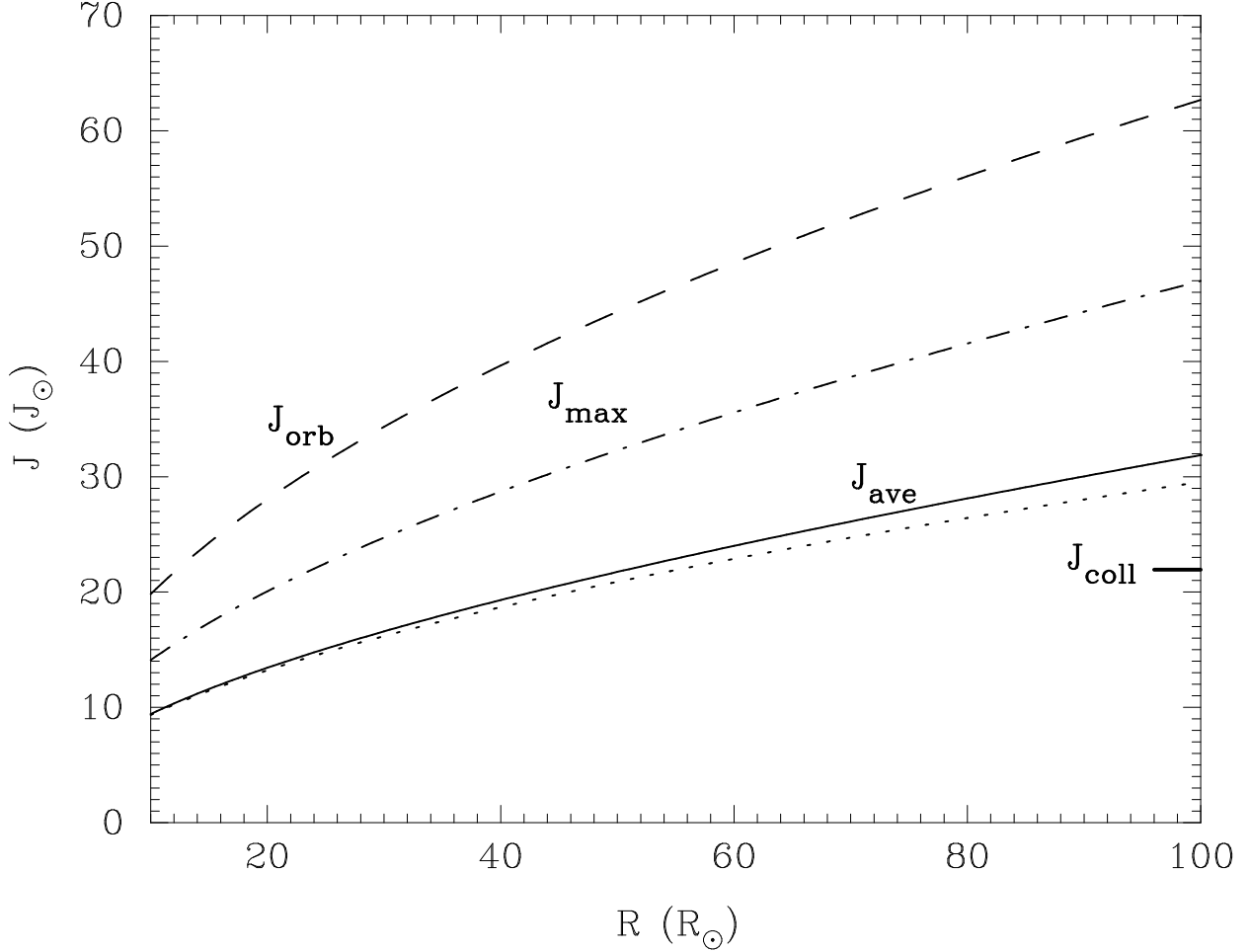}
\caption{
%{\bf FIGURE CAPTIONS}
%\noindent {\bf Figure 1:}
Planet-deposited angular momentum, in
units of the solar angular momentum $1.7 \times 10^{48} \g \cm^2 \s^{-1}$,
as a function of the RGB stellar radius at the time of collision.
Plotted are the average angular momentum deposited by FFPs,
$J_{\rm ave}$: ``accurate'' expression  (eq. 11; solid line),
and an approximate expression (eq. 12;
dotted line); maximum angular momentum deposited by a FFP,
$J_{\rm max}$ (eq. 13; dot-dashed line); and angular momentum deposited
by an orbiting planet, $J_{\rm orb}$, which is engulfed
due to tidal forces when the parent star expands to $\sim 1/4$ of
the orbital separation (eq. 14; dashed line).
All calculations assume a planetary mass equal to that of Jupiter,
$M = 0.85 M_\odot$, and $v_0 = 20$ km s$^{-1}$.
 Also marked ($J_{\rm coll}$) the average deposited angular momentum
over all collisions in all RGB stars, assuming the RGB terminates
at $R=100 R_\odot$ and $m_p=M_{\rm J}$.
Jupiter's orbital angular momentum is $1.93\times 10^{50}
\g\cm^2\s^{-1}$, about 100 times that of the sun, so
values in the plot are approximately the {\em percentage}
of Jupiter's orbital angular momentum.
}
\end{figure}
\end{document}